**O. V. KOMPANIIETS**, junior researcher at the Main Astronomical Observatory of the NAS of Ukraine,
Post-graduate student at the Institute of Physics of the NAS of Ukraine
https://orcid.org/0000-0002-8184-6520
E-mail: kompaniets@mao.kiev.ua

Main Astronomical Observatory of the National Academy of Sciences of Ukraine
27 Akademik Zabolotnyi Str., Kyiv, 03143, Ukraine

# MULTIWAVELENGTH PROPERTIES OF THE LOW-REDSHIFT ISOLATED GALAXIES WITH ACTIVE NUCLEI MODELLED WITH CIGALE

*Using the CIGALE software, we present the preliminary results of a multiwavelength analysis of eighteen low-redshift isolated galaxies with active nuclei (isolated AGNs). This sample was formed by cross-matching the 2MIG isolated AGNs sample with the SDSS DR9 catalog. The host galaxies of this sample have not undergone a merger for at least three billion years, making them a unique laboratory for studying interactions between various astrophysical processes without the complicating factors of merging with other galaxies or the effects of a denser environment. In addition, the study of isolated AGNs can provide valuable information about the evolution and activity of galaxies in the broader context of the distribution of large-scale structures of the Universe. First, we seek to understand how the environment affects the physical processes involved in the accretion of matter onto supermassive black holes in these galaxies. Secondly, to what extent do processes of star formation or degeneration of nuclei activity continue the evolution of these galaxies? Third, how does the localization of isolated AGNs in voids or filaments of a large-scale structure determine the properties of this environment at the low redshifts?*

*Using observable fluxes from UV to the radio ranges from archival databases of space-born and ground-based observatories (GALEX, SDSS, 2MASS, Spitzer, Hershel, IRAS, WISE, VLA), we estimated the contribution from the emission of an active nucleus to the galaxy's total emission, the stellar mass, and the star formation rate. The mass of the stellar component falls from $10^{10}\,M_{Sun}$ and $10^{11}\,M_{Sun}$. The star formation rate for most galaxies (except UGC 10120) does not exceed 3 $M_{Sun}$ per year. The best SED fittings (with $\chi^2$ values less than 5) are obtained for the galaxies CGCG248-019 ($\chi^2 = 1.6$), CGCG179-005 ($\chi^2 = 1.6$), CGCG243-024 ($\chi^2 = 2.6$), IC0009 ($\chi^2 = 2.8$), MCG+09-25-022 ($\chi^2 = 3.1$), UGC10244 ($\chi^2 = 4.1$).*

*Keywords: galaxies, isolated galaxies, active galaxy nuclei, stellar mass, star-formation rate; objects: CGCG248-019, CGCG179-005, CGCG243-024, IC0009, MCG+09-25-022, UGC10244.*

## 1. INTRODUCTION

The isolated galaxies are important components of the large-scale structure (LSS) of the Universe. Their location in the lower environment allows one to study the physical properties (star-formation rate, nuclear activity, morphological and multiwavelength features, mass distribution, interstellar medium, satellite galaxies at the outskirts of their host haloes, past merging, etc.) without the influence of significant neighboring galaxies as comparing with galaxies in the tightly populated groups or clusters (see, the pioneering works by Karachentseva [15—17] as well as other articles [6, 12, 14, 18, 27, 28, 30, 39, 40, 43], e.g., in frame of the collaborative AMIGA project [33, 34, 37, 47]. For example, a comprehensive analysis of the 2MASS Isolated Galaxy catalog (2MIG) in the near-infrared and optical ranges [28] has demonstrated a significant impact of the environment on the color parameters of isolated galaxies. In this sense, isolated galaxies with active nuclei (isolated AGNs) are excellent laboratories for studying active and other processes regulated by only intrinsic factors and analyzing the presence/absence of feedback from an active nucleus [10, 31].

We consider the sample of 61 isolated AGNs at the redshift $z < 0.05$, which was formed by cross-matching the 2MIG catalog [19] with the Catalog of quasars and AGNs [48], where the restriction was used for stellar magnitude $K_s < 12.0$ mag and radial velocity $V_r < 15000$ km/s [31]. We used available observational data from different ground-based and space-born observatories to obtain multiwavelength properties of 2MIG isolated AGNs and to estimate the general properties of these galaxies.

Multiwavelength properties of several galaxies from this sample of low redshift galaxies were already provided, including properties in the X-ray spectral range [22, 26, 50, 52]. It was noted that all these objects are very faint in X-ray in comparison with the AGNs in a more dense environment [1, 9, 11, 23, 25, 35, 49]. A part of these galaxies has a reflection component in X-ray spectra with different reflection fractions in addition to the primary power-law continuum [38, 45]. In the latest research [32], the radio properties of 61 isolated AGNs were analyzed: the typical spectral flux densities at 1.4 GHz are between 3 and 20 mJy, so these galaxies are also faint in the radio band; however, two galaxies, PGC35009 and NGC6951, displayed higher-than-average flux densities in the 50 to 200 mJy range; in contrast, two galaxies, ESO483-009 and ESO097-013, exhibited substantial spectral flux densities of 352 and 1200 mJy, respectively, while flux densities for 10 isolated AGNs were less than 3 mJy. At the same time, an analysis of the AMIGA sample of isolated galaxies revealed that galaxies within this sample were also notably radio-quiet [26]. Another feature of these isolated AGNs is the lower masses of supermassive black holes (SMBH) [21, 46].

## 2. SOFTWARE & DATA PREPARATION

**2.1. *CIGALE*.** The sophisticated software tools have become indispensable in galaxy evolution research for unraveling the intricate processes governing the formation and evolution of galaxies. In this research, we used the CIGALE — a Python Code Investigating GALaxy Emission. This software has been developed by Boquien et al. [3] to study the evolution of galaxies by comparing the modeled galaxy spectral energy distributions (SEDs) to observed ones from the far ultraviolet to the far infrared and radio. Recently, a new X-ray module for CIGALE, allowing it to fit SEDs from the X-ray to infrared (IR), was developed to improve the AGN fitting [51].

To compute spectral models, CIGALE constructs complex stellar populations from simple stellar populations combined with highly flexible star formation histories, calculates the emission of gas ionized by massive stars, and attenuates both stars and ionized gas with a highly flexible attenuation curve. Based on the principle of energy balance, the absorbed energy is then re-emitted by the dust in the mid- and far-infrared regions, and thermal and non-thermal components are also included, extending the spectrum far into the radio range. A large grid of models is then fitted to the data, and the physical properties are estimated through probability distribution analysis[1]. The CIGALE software flexibility allows us to work with the data of various observational sky surveys, catalogs, and archives (see review in [44]), to develop spectroscopic galaxy classification using unsupervised technique [8], to select candidates into high-redshift AGNs using the Early Release Observations data from JWST [13], to identify a new type of infrared-bright dust-obscured galaxies (overweight DOGs), which then become visible quasars [36], to evaluate a role of environment in suppressing the star formation process and following morphological transformation from late-type spirals to early-type galaxies in compact galaxy groups [2], etc. For example, a large public catalog of about 1.9 million galaxies from an eBOSS (Extended Baryon Oscillation Spectroscopic Survey) at redshifts $z = 0{-}1.5$ [29], which includes SEDs from the Sloan Digital Sky Survey (SDSS), *ugriz* photometry, and the available WISE mid-infrared photometry modeled with CIGALE, will provide yet more tasks to study the dependence of galactic outflows on host galaxy physical properties.

---

[1] https://cigale.lam.fr/documentation/

*Table 1.* **Parameters for converting stellar magnitudes into fluxes for the SDSS filters**

| Filter | $b$ | Zero-flux magnitude $[m(f/f_0 = 0)]$ | $m(f/f_0 = I_0)$ |
|--------|-----|------------------------------|------------------|
| $u$ | $1.4 \times 10^{-10}$ | 24.63 | 22.12 |
| $g$ | $0.9 \times 10^{-10}$ | 25.11 | 22.6 |
| $r$ | $1.2 \times 10^{-10}$ | 24.8 | 22.29 |
| $i$ | $1.8 \times 10^{-10}$ | 24.36 | 21.85 |
| $z$ | $7.4 \times 10^{-10}$ | 22.83 | 20.32 |

*Table 2.* **Input parameters for the SED fitting of isolated AGNs**

| Parameter | Values |
|-----------|--------|
| *Star formation history* | |
| E-folding time of the main stellar population in Myr | 500, 1000, 2500, 6000 |
| Age of the main stellar population in the galaxy in Myr | 5000, 8000, 13000 |
| *Stellar populations* | |
| Initial mass function | Chabrier |
| Metallicity | 0.008, 0.02 |
| Age in Myr of the separation between the young and the old star populations | 100, 500, 1000, 5000, 8000 |
| *Nebular emission* | |
| Ionisation parameter | –1.0, –2.0, –4.0 |
| Gas Metallicity | 0.002, 0.02 |
| *Dust attenuation* | |
| V-band attenuation in the interstellar medium | 0.15, 0.45, 0.75, 1.05, 1.5, 2.1, 4.0 |
| Power law slope of the attenuation in the ISM | –0.7 |
| Power law slope of the attenuation in the birth clouds | –1.3 |
| *Dust emission* | |
| Alpha slope | 0.5, 2.0, 4.0 |
| *Synchrotron radio emission* | |
| The slope of the power-law synchrotron emission related to star formation | 0.8 |
| The slope of the power-law AGN radio emission (assumed isotropic) | 0.7 |
| *AGN* | |
| The ratio of the maximum to minimum radii of the dust torus | 10.0, 60.0 |
| Opening angle | 60.0 |
| The angle between the equatorial axis and line of sight | 0.001, 89.99 |
| $E(B - V)$ for the extinction in the polar direction in magnitudes | 0.25, 0.5, 1.0 |
| Temperature of the polar dust in K | 25 |

**2.2. *Data preparation.*** An initial sub-sample of isolated AGNs for multiwavelength analysis was obtained by cross-matching the studied sample of 61 isolated AGNs with the morphological SDSS DR9 catalog of galaxies at $z < 0.1$ [41, 42] classified by the machine learning methods with photometry-based [46] and image-based approaches [20]. As a result, only 18 isolated AGNs were selected for further analysis. Available observed stellar magnitudes of galaxies in the UV range were obtained by cross-matching the resulting sample with the GALEX, infrared — 2MASS, and WISE catalogs.

There is a special procedure to convert the observed magnitudes into fluxes in mJ for further analysis. The SDSS catalog contains stellar magnitudes in 5 photometry ranges ($u, r, g, i, z$) and their errors. The value of the observed magnitude is related to the flux through an inverse hyperbolic sine and is given as follows[2]:

$$m = -2.5 / \ln(10) \times [asinh((f / f_0)/(2b)) + \ln(b)]. \quad (1)$$

Here, $m$ is the magnitude in the $u, r, g, i, z$ filter, $f$ is the flux, $f_0$ is the flux for zero magnitude $m_0$, $b$ — coefficient. These values differ for each filter $u, r, g, i, z$. They are shown in Table 1.

After solving equation (1), we get the following relationship between the flux and the magnitude:

$$f = \sinh\left(-0.921m - 2\times10^3 \, bf_0 \log(b)\right), \quad (2)$$

where $f_0 = 3631$ Jy.

The observed magnitudes were converted from the 2MASS, GALEX, and WISE catalogs to the Jy fluxes using the usual Pogson formula, considering the zero magnitudes and the corresponding flux value for each catalog. However, the data for isolated AGNs are absent in the archives of the Spitzer, Hershel, and IRAS space observatories. So, we provided an individual search of observational data from these space missions and for radio ranges in the NED database.

## 3. SED MODEL & PRELIMINARY RESULTS

For the first step in our analysis, we ran different tests to choose a baseline model to describe emission from the ultraviolet to radio ranges. One can choose a combination of physical models that describe emis-sions from different components of galaxies. Input parameters for the SED fitting are given in Table 2. Among these physical models are as follows:

• sfhdelayed — delayed SFH with optional exponential burst,

• bc03 — stellar population synthesis [4],

• nebular — continuum and line nebular emission,

• dustatt_modified_CF00 Calzetti 2000 attenuation law [5],

• dale2014 — dust emission templates [7],

• fritz2006 — AGN models [11],

• radio — galaxy synchrotron emission and AGN,

• redshifting.

---

[2] https://www.sdss.org/dr12/algorithms/magnitudes/#asinh

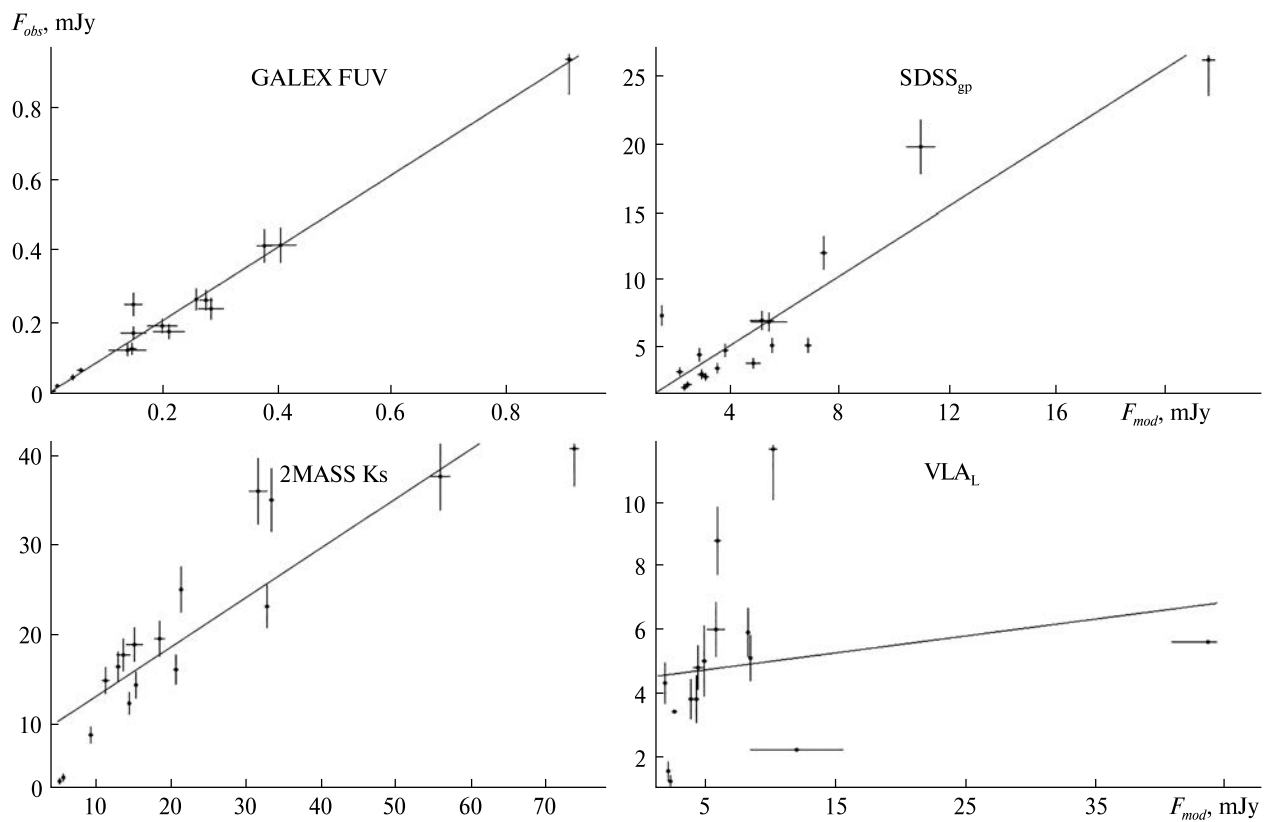

**Figure 1.** Observed fluxes $F_{obs}$ in different bands vs model fluxes $F_{mod}$ of the isolated low-redshift AGNs

All the obtained physical properties (star-formation rate (SFR), stellar mass, AGN fraction, etc.) are model-dependent. Here, we present preliminary results of one of the possible model combinations for describing multiwavelength emission from isolated AGNs.

All the studied galaxies are spiral but with different spectral activity types of their AGN (Table 3, col-umn 3), which plays an essential role in observable SED. In our model, we consider the presence of AGN using model [50], where the angle between the equatorial axis and line of sight was set to 90° for type Sy1 and 0° for type Sy2. This model describes well the SED in ultraviolet and optical ranges, see Figure 1, where the vertical axis shows the observed fluxes in different ranges, and the horizontal axis shows the simulated fluxes. The best correlation coefficients of 0.99 were obtained for GALEX FUV, 0.97 for SDSS_z, and 0.93 for SDSS_g. In the infrared range, the coefficient 0.88 for 2MASS Ks is the best match. The far-infrared and radio components require the use of a different model, so the observed magnitudes are well described for only a few galaxies, and the overall correlation coefficient is 0.22 for VLA_L data.

A good fit in the first approximation was obtained for 6 galaxies — CGCG 179-005, IC0009, CGCG 243-024, MCG +09-25-022, UGC 10244, UGC 06398 (Figure 2). The obtained value of the AGN fraction

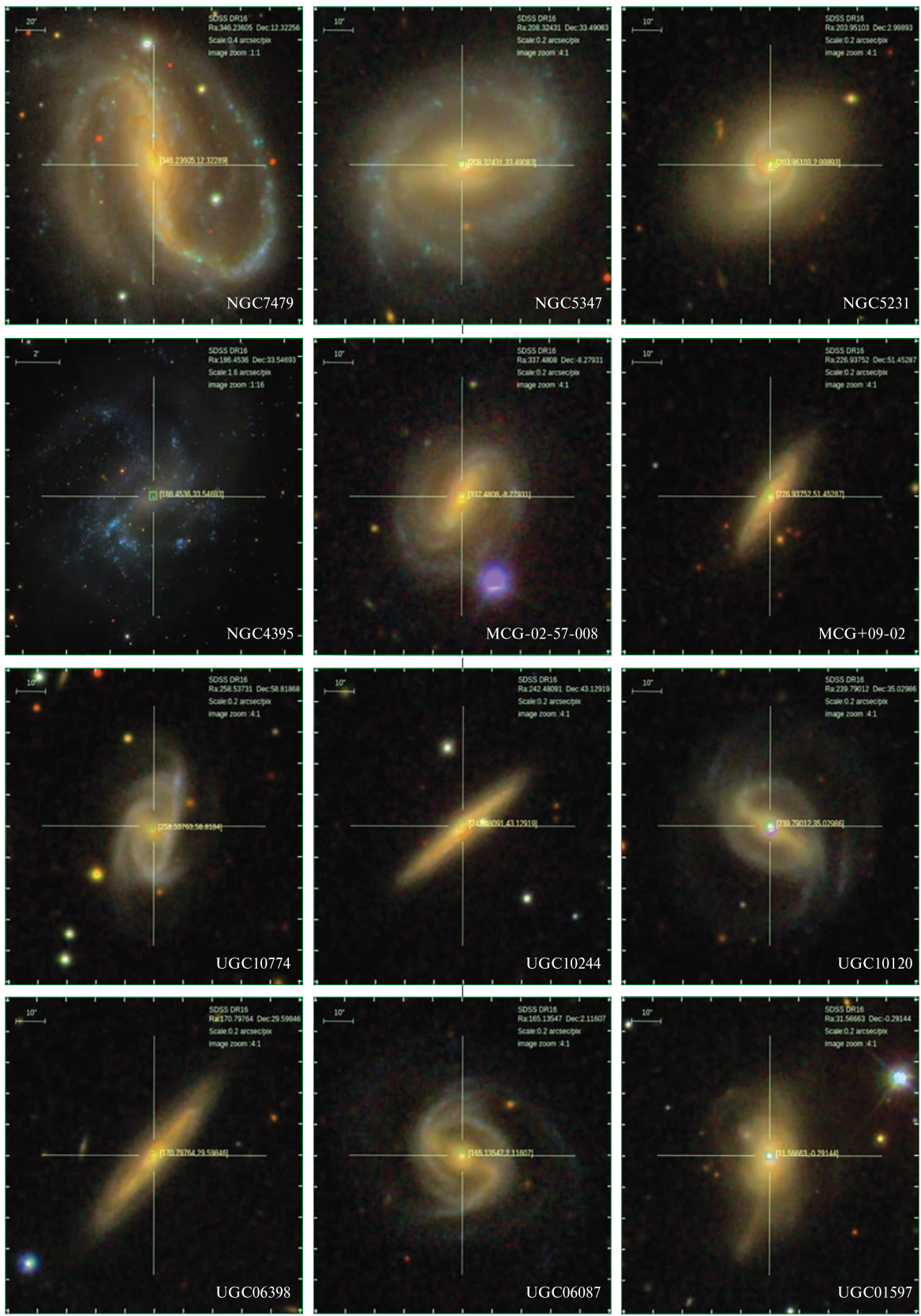

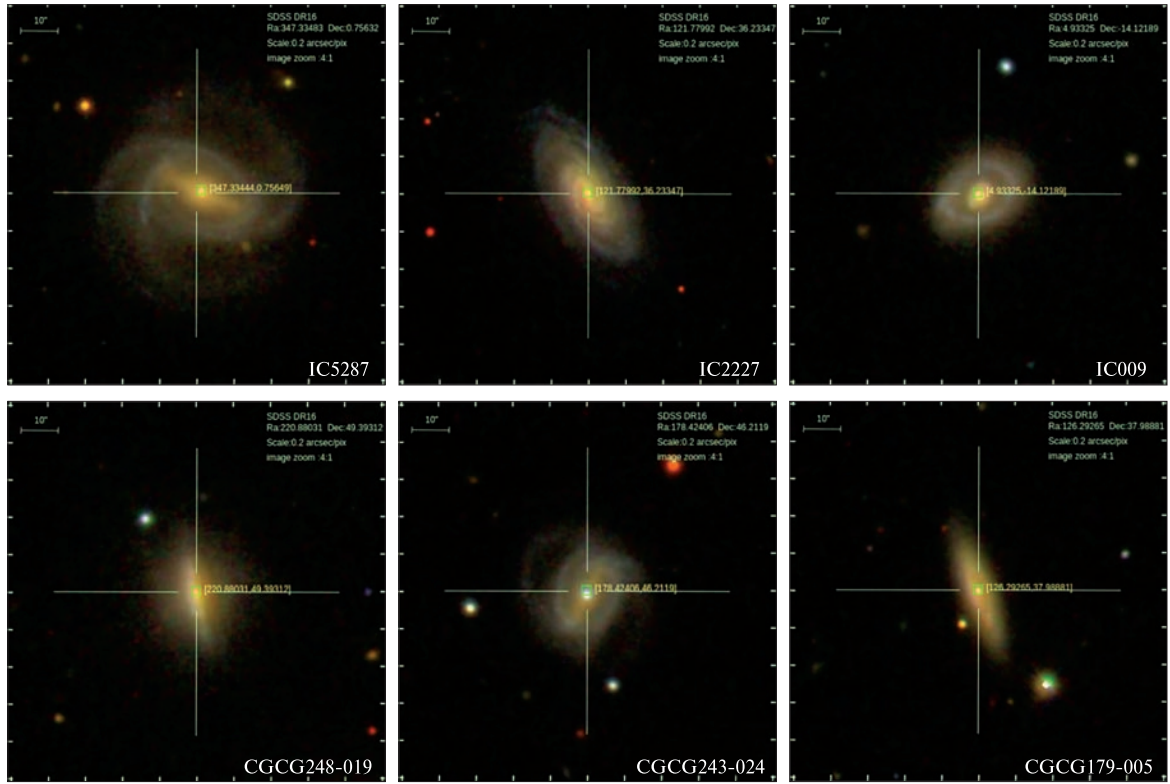

***Figure 2.*** SDSS images [https://skyserver.sdss.org/dr16] of 18 2MIG isolated AGNs, which were modeled with CIGALE

*Table 3.* **General properties of 2MIG isolated AGNs**

| Name | Morph. type | Activity type | Features | Stellar mass, $10^{10}M_{Sun}$ | SFR, $M_{Sun}\,y^{-1}$ | $\chi^2$ | $F_{15-150\,keV}$ $10^{-15}\,W/m^2$ |
|---|---|---|---|---|---|---|---|
| UGC10774 | SABb | NLAGN | Bar, ring | 0.9 | 1.11 | 12 | — |
| UGC10244 | Sbc | LINER | SN Ia | 4.6 | 0.24 | 4.1 | |
| UGC10120 | SB(r)b | Sy 1n | Bar, ring | 4.5 | 5.42 | 11 | |
| UGC06398 | Sbc | Sy 2 | Ring | 9.6 | 0.16 | 6.4 | |
| UGC06087 | SBb | ? | Bar, ring | 5.4 | 1.24 | 15 | |
| UGC01597 | S0 | Sy 1.9 | — | 17.5 | 0.89 | 22 | 24.27 |
| NGC7479 | SB(s)c | Sy 2 | Bar, SN 1b, SN | 1.2 | 0.02 | 47 | 14.69 |
| NGC5347 | SB(rs)ab | Sy 2 | Bar, ring | 0.9 | 0.58 | 13 | |
| NGC5231 | SBa | Sy 1 | Bar | 5.5 | 1.28 | 9.4 | 7.7 |
| NGC4395 | SA(s)m | Sy 1.8 / Linear | Ring | 0.001 (unnfit) | 0.0007 (unfit) | 37 | 27.53 |
| MCG-02-57-008 | Sc | | — | 6.3 | 0.11 | 16 | |
| MCG+09-25-022 | Sa | Sy 1.0 | Ring | 11.1 | 2.56 | 3.1 | |
| IC5287 | SBb | Sy 1.2 | Bar | 4.9 | 1.14 | 5.5 | |
| IC2227 | SBa | Sy 2 | Bar | 12.4 | 2.82 | 13 | |
| IC0009 | Sa | Sy 2 | Ring | 4.7 | 3.03 | 2.8 | |
| CGCG248-019 | SBab | Candidate BLAGN | Bar | 3.1 | 1.93 | 1.6 | |
| CGCG243-024 | SBb | Sy1 | Bar, ring | 1.1 | 1.21 | 2.6 | |
| CGCG179-005 | Sbc | Candidate BLAGN | — | 0.8 | 0.54 | 1.6 | |

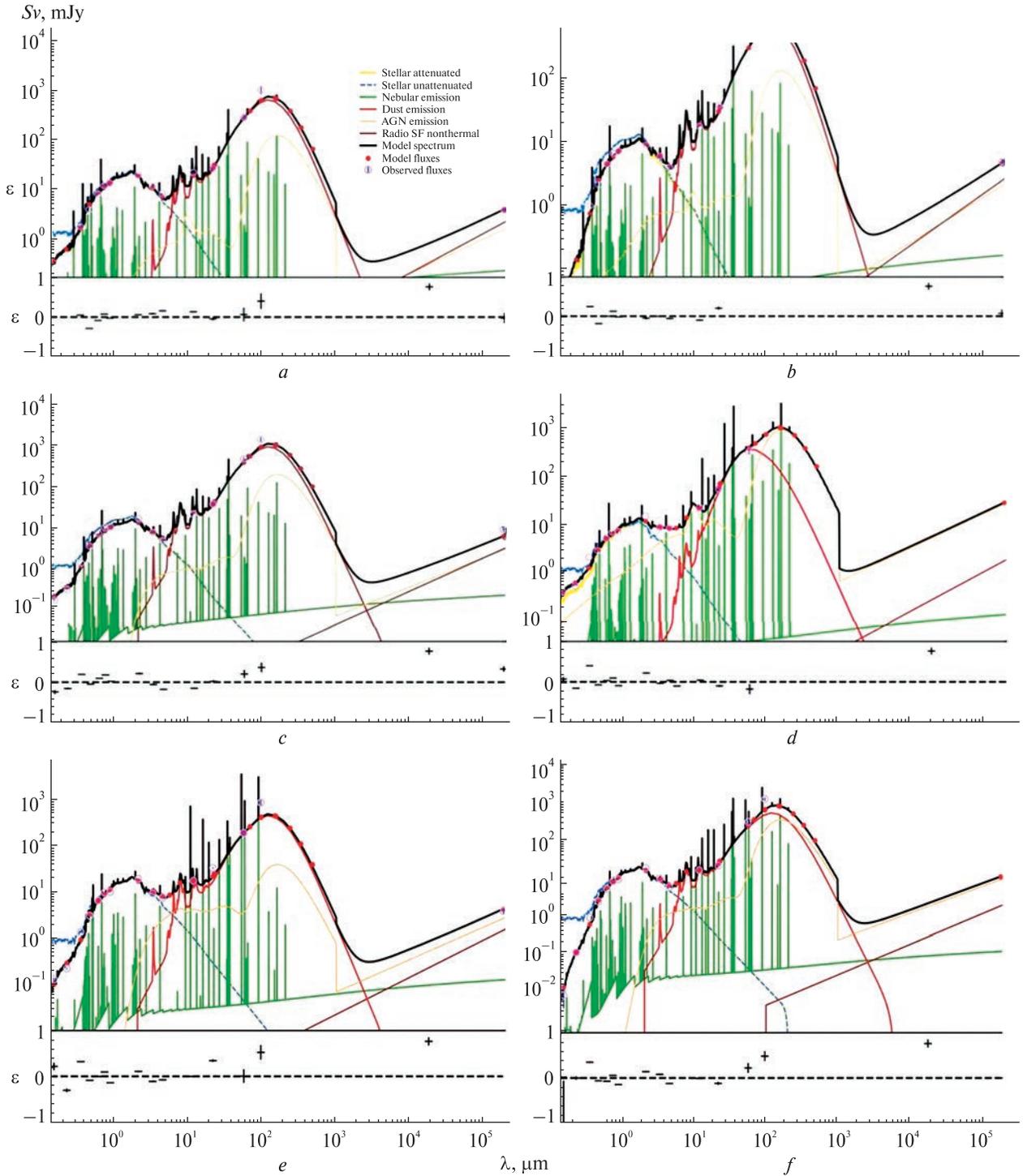

**Figure 3.** The best-fit results for SEDs fitting for several isolated AGNs: $a$ — CGCG248-019, $z = 0.0302$, $\chi^2 = 1.6$; $b$ — CGCG179-005, $z = 0.0214$, $\chi^2 = 1.6$; $c$ — IC0009, $z = 0.0421$, $\chi^2 = 2.8$; $d$ — CGCG243-024, $z = 0.0243$, $\chi^2 = 2.2$; $e$ — MCG+09-25-022, $z = 0.0459$, $\chi^2 = 3.3$; $f$ — UCG10244, $z = 0.0325$, $\chi^2 = 2.5$

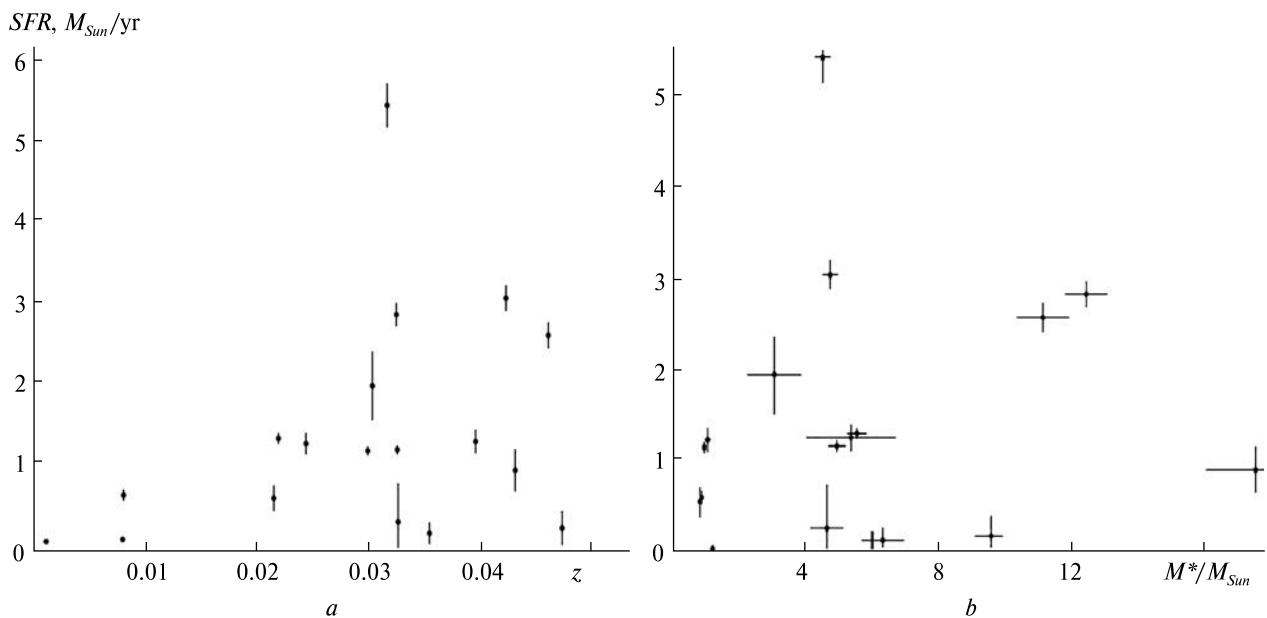

**Figure 4.** The *SFR* distribution: *a* — for studied isolated AGNs with redshift; *b* — the distribution of estimated *SFR* with stellar mass of studied isolated AGNs

is 0.1 for 15 objects and 0.5 for three objects (UGC 10244, MCG-02-57-008, and CGCG 243-024). The evaluated stellar masses and star-formation rates are presented in Table 3 (columns 6 and 7, respectively).

The studied 18 AGNs have different morphological types and features with a bar, without a bar, with a bar and a ring (Figure 2, Table 3, columns 3 and 5), and types of nuclear activity as Seyferts 1 and 2, Liners (Table 3, column 4). For this reason, we cannot distinguish certain morphological features common to objects which the basic model will describe well. However, galaxies for which the chosen model does not give the desired result also have emission in the hard X-ray as evidenced by the data from the 150-month Swift/BAT survey [24], so we suggest that the actual contribution from AGN is much more significant for some galaxies than is estimated in our preliminary model.

## 4. DISCUSSION

We present preliminary results of multiwavelength properties of 18 isolated AGNs modeled with the CIGALE software. Analysis of the emission in a broad range by the baseline model showed that it describes the spectral energy distribution for 6 galaxies quite well (see Table 3, column 8 for $\chi^2 < 5.0$). The best SED fittings of CGCG248-019 ($\chi^2 = 1.6$), CGCG179-

005 ($\chi^2 = 1.6$), CGCG243-024 ($\chi^2 = 2.6$), IC0009 ($\chi^2 = 2.8$), MCG+09-25-022 ($\chi^2 = 3.1$), UGC10244 ($\chi^2 = 4.1$) are presented in Figure 3.

The contribution to the emission from the active nucleus for 15 galaxies is estimated at 0.1 %. It is consistent with our conclusion that the activity of the nucleus of the most isolated galaxies is faint. However, this assumption is not valid for a few objects, in particular, UGC01597, NGC7479, NC5231, and NGC4395, since they have hard X-ray emission according to the data from the Swift/BAT catalog, and their AGN luminosity varies from $10^{42}$ to $10^{44}$ erg/s (Table 3, column 9).

Whereas the quantification of stellar mass ($M^*$) and the assessment of star formation rates (SFR) represent pivotal parameters in the characterization of galaxies, accurate measuring of these fundamental properties is paramount in elucidating the present state of galaxies, their history, and future evolution. To estimate these physical quantities, we tested different models for describing the star formation history (SFH) of the stellar population. As a result, a "delayed" SFH is chosen, where the evolution of the star-forming rate is described as follows :

$$\text{SFR}(t) \propto \frac{t}{\tau^2} \times \exp(-t/\tau) \text{ for } 0 \leq t \leq t_0,$$

here $t_o$ — the age of the onset of star formation, and $\tau$ is the time at which the SFR peaks. After peaking at $t = \tau$, it smoothly decreases.

For the stellar population description, the bc03 module was chosen, where a single stellar population (SSPs) was used [4]. SSP library is available for a broad range of metallicities (0.0001, 0.0004, 0.004, 0.008, 0.02, and 0.05). To compute the spectrum of the composite stellar populations, CIGALE calculates the dot product of the SFH with the grid containing the evolution of the spectrum of an SSP with steps of 1 Myr [3].

As for the relatively accurate description of the ultraviolet and optical components of the SED concerning the stellar mass and star formation rate, we conclude as follows. The estimated mass of the stellar component falls from $10^{10}$ $M_{Sun}$ and $10^{11}$ $M_{Sun}$ (Table 3, column 6). The star formation rate for most galaxies (except UGC 10120) does not exceed 3 $M_{Sun}$ per year (Figure 4, *a*; Table 3, column 7), indicating the absence of active star formation. Separately, it is worth noting that the selected model does not describe the galaxy NGC 4395. It may be related to a different ratio between the old and young stellar populations (see Figure 2, SDSS image of NGC 4395). The SFR distribution shows a trend towards an increase in the star formation rate with increasing redshift (Figure 4, *a*). We note that SFR for 1616 isolated galaxies selected from the Two-Micron All-Sky Survey were obtained in [28] based only on the GALEX data. The estimated star-forming rate per stellar mass is presented in Figure 4, *b*: due to the small statistics, the conclusion is premature.

The best-fit results for SEDs fitting for other isolated AGNs, as well as their physical properties, will be analyzed with other models.

**Acknowledgements.** *I express my gratitude to Prof. Iryna Vavilova for useful discussions, comments, and remarks. The work was supported by the Target scientific project of the National Academy of Sciences of Ukraine "Multiwavelength properties of isolated galaxies with active nuclei" (0123U102380), a grant for Research works of young scientists of the National Academy of Sciences of Ukraine (2023—2024, 0123U103122), and scholarship of the President of Ukraine for young scientists.*

*О. В. Компанієць*, мол. наук. співроб., аспірантка Інституту фізики НАН України
https://orcid.org/0000-0002-8184-6520
E-mail: kompaniets@mao.kiev.ua

Головна астрономічна обсерваторія Національної академії наук України
вул. Академіка Заболотного 27, Київ, Україна, 03143

БАГАТОХВИЛЬОВІ ВЛАСТИВОСТІ БЛИЗЬКИХ ІЗОЛЬОВАНИХ
ГАЛАКТИК ІЗ АКТИВНИМИ ЯДРАМИ: CIGALE-МОДЕЛЮВАННЯ

Представлено попередні результати багатохвильових властивостей вісімнадцяти ізольованих галактик з активними ядрами, змодельованих у програмному середовищі CIGALE. Вибірку галактик було сформовано методом крос-кореляції вибірки ізольованих 2MIG галактик з активними ядрами (АЯГ) з каталогом SDSS DR9. Материнські галактики цієї вибірки не зазнавали злиття протягом щонайменше трьох мільярдів років, що робить їх унікальною лабораторією для для вивчення взаємодії між різними астрофізичними процесами без факторів, що ускладнюють взаємодію з іншими галактиками, або ефектів середовища щільного скупчення. Крім того, дослідження ізольованих галактик з АЯГ може дати цінну інформацію про еволюцію та активність галактик у ширшому контексті розподілу великомасштабних структур Всесвіту. По-перше, ми прагнемо зрозуміти, як оточення впливає на фізичні процеси, пов'язані з акрецією речовини на надмасивні чорні діри в цих галактиках. По-друге, якою мірою процеси зореутворення чи дегенерації активності ядра продовжують еволюцію цих галактик. По-третє, яким чином локалізація ізольованих АЯГ у войдах чи філаментах великомасштабної структури визначає властивості цього середовища на малих червоних зміщеннях.

Використовуючи спостережувані потоки від УФ- до радіодіапазонів з архівних баз даних (GALEX, SDSS, 2MASS, Spitzer, Hershel, IRAS, WISE, VLA), ми оцінили внесок випромінювання активного ядра в загальне випромінювання галактики, зоряну масу і швидкість зореутворення. Маса зоряного компонента для проаналізованих галактик лежить у межах від $10^{10}$ $M_{Sun}$ до $10^{11}$ $M_{Sun}$. Швидкість зореутворення для більшості галактик (крім UGC 10120) не перевищує 3 $M_{Sun}$ за рік. Найкращі моделі спектрального енергетичного розподілу (з $\chi^2 < 5$) отримано для галактик CGCG248-019 CGCG248-019 ($\chi^2 = 1.6$), CGCG179-005 ($\chi^2 = 1.6$), CGCG243-024 ($\chi^2 = 2.6$), IC0009 ($\chi^2 = 2.8$), MCG+09-25-022 ($\chi^2 = 3.1$), UGC10244 ($\chi^2 = 4.1$).

*Ключові слова*: галактики, ізольовані галактики, активні ядра галактик, зоряна маса, темп зореутворення; об'єкти: CGCG248-019, CGCG179-005, CGCG243-024, IC0009, MCG+09-25-022, UGC10244.